# Neutron structural studies on the superconducting $(Nd_{1-x}Ca_x)(Ba_{1.6}La_{0.4})Cu_3O_z$ system

Amish G. Joshi[a,*], R.G. Kulkarni[b], W.B. Yelon[c], Ram Prasad[d], M.R. Gonal[d]

[a.] *National Physical Laboratory, Dr. K.S. Krishnan Road, New Delhi – 110 012, India*
[b.] *Department of Physics, Saurashtra University, University Road, Rajkot – 360 005, India*
[c.] *Graduate Center for Material Research, University of Missouri, Rolla, MO – 65409, USA*
[d.] *Metallurgy Division, Bhabha Atomic Research Center, Trombay, Mumbai – 400 085, India*

## Abstract:

We have investigated the influence of Ca ions substitution on the structural and superconducting properties of $(Nd_{1-x}Ca_x)(Ba_{1.6}La_{0.4})Cu_3O_z$ system. Magnetization, X-ray diffraction and neutron diffraction studies have been carried out on a series of compounds with $x$ = 0.0 to 0.6. The superconducting transition temperature $T_c$, determined from magnetization measurements increases with increasing $Ca^{2+}$ substitution. Neutron diffraction studies reveal that these compounds crystallize in a tetragonal structure (space group $P4/mmm$). A detailed analysis of the neutron diffraction data reveals that Ca and La ions are intermixed at the nominal Ba and Nd sites. While a major fraction of Ca ions occupy the usual Nd site, a small fraction occupies the Ba site. Consequently, the corresponding amount of La substitutes at the nominal Nd site. The intermixing of Ca and La sites randomizes the chain site oxygens leading to a tetragonal structure despite an oxygen content close to 7.0 for all the Ca doped samples. Further increase of Ca content lead to change in its co-ordination from six-fold to eight-fold at $x \geq 0.4$.

---

*Corresponding author: amish@nplindia.org

# 1. Introduction

The normal state transport properties of oxide superconductors are unusual and systematic studies of them are therefore important for understanding high $T_c$ superconductivity. $YBa_2Cu_3O_{7-\delta}$ (Y123) superconductor has been extensively studied; replacing Y by R (R = Rare Earth) does not affect superconductivity except when R is Ce, Tb or Pr. The hole concentration, $p$, is an important parameter in the p-type ceramic high $T_c$ materials. The hole concentration can be varied by changing the concentration of substituents at different crystallographic sites (i.e. Y, Ba or Cu), or by the ordering of oxygens in basal plane. It is well known that in $R_{1+x}Ba_{2-x}Cu_3O_{7-\delta}$ the light rare earth element (LR = La to Nd) can occupy the Ba site as well as the R site. For LR elements the ionic radii approach that of Ba [1, 2], allowing for a large degree of substitution by R for Ba without forming second phase, which results in oxygen defects on the anti chain site. Nd has an ionic radius closest to that of Ba and, therefore, exhibits the greatest amount of solubility with Ba ($x \cong 0.8$) [3]. When trivalent LR occupies the divalent Ba site, charge balance requires ½ oxygen for each R is incorporated into the structure. In the fully oxygenated orthorhombic structure the extra oxygen occupies the anti-chain site [O(5)] [4-6]. Since the basal plane acts as a charge reservoir, the extra oxygen results in modification of the charge distribution in the structure. Superconductivity is suppressed due to the localization of holes [6-8]. Accompanying the $LR^{3+}$ substitution on the $Ba^{2+}$ site is a change in the crystal structure from orthorhombic to tetragonal (O-T) and an increase in the flatness of the $CuO_2$ plane [9]. A comparison of substituting magnetic ions (such as $Eu^{3+}$, $Pr^{3+}$, $Nd^{3+}$) [7,9,10] and non-magnetic ions (such as $La^{3+}$) [10-13] at Ba site leads to the conclusion that besides magnetic pair breaking, there is a considerable contribution to the $T_c$ originating from other effects like defect scattering, carrier localization or hole filling.

By studying La substituted Nd123 compounds we have shown that increasing La concentration in Nd(Ba$_{2-y}$La$_y$)Cu$_3$O$_z$ ($0 \leq y \leq 0.5$) causes a reduction in $T_c$ due to hole filling, and a structural phase transition (O-T) is observed at $y = 0.3$ [11]. A similar structural transformation is observed with nonmagnetic La$^{3+}$ at Ba$^{2+}$, with magnetic rare earth R (R = Eu, Ho, Dy) substitutions [10, 12, 13]. This $T_c$ is expected to revive when divalent Ca is substituted for trivalent rare earth R. When Ca is incorporated in (Nd$_{0.94}$Ca$_{0.06}$)Ba$_2$Cu$_3$O$_{7-\delta}$ the sample is observed to be more metallic at 8 GPa and 523 K [14]. On the other hand, thermoelectric power measurements have shown that La doping in Nd$_{1-x}$La$_x$Ba$_2$Cu$_3$O$_y$ causes a decrease in hole concentration smaller than that of Ca doping in Nd$_{1-x}$Ca$_x$Ba$_2$Cu$_3$O$_y$, which suggests that some La (~26%) would enter in to the Ba site with the generation of additional BaCuO$_{2+v}$ phase [15]. Our present efforts are to understand the effect of simultaneous substitution of Ca and La at the Nd and Ba sites, respectively, on structure and superconductivity of (Nd$_{1-x}$Ca$_x$)(Ba$_{1.6}$La$_{0.4}$)Cu$_3$O$_z$ system. If all the La atoms occupy the Ba site, the oxygen content would have to increase to maintain charge balance. As Ca (1.12 Å, in 8 fold coordination) is slightly larger than Nd (0.995 Å), there should be a preference for Ca to occupy the Ba-site rather than the Nd site, which is solely based on lattice strain arguments. However, this preference is not observed, and is expected to be due to some inter site mixing of Ca and La at the Ba and Nd sites, respectively. For this reason neutron diffraction studies were performed to determine the relative site occupations of the Ca and La ions. A systematic investigations of neutron diffraction and magnetization measurements of compounds having the stoichiometric compositions (Nd$_{1-x}$Ca$_x$)(Ba$_{1.6}$La$_{0.4}$)Cu$_3$O$_z$ with $x = 0.0$ to 0.6 have been studied in detail. The interrelationship between the superconducting transition temperature ($T_c$) and structural changes has also been discussed in the context of the carrier concentration.

## 2. Experimental details

The $(Nd_{1-x}Ca_x)(Ba_{1.6}La_{0.4})Cu_3O_z$, series of compounds with $0 \leq x \leq 0.6$, were synthesized via a solid state reaction, the details of which are given in Ref. [14]. Powder X-ray diffraction patterns of the compounds were obtained in the $2\theta$ range $20^\circ \leq \theta \leq 80^\circ$ using Cu $K_\alpha$ radiation.

Room temperature neutron diffraction patterns, were obtained at the University of Missouri Research Reactor Facility using a position sensitive detector (PSD) diffractometer. This diffractometer is equipped with a Si [511] monochromatic, horizontally curved with about 11-12 m radius and vertically segmented with 1.5 m radius and with 9 asymmetrically cut blades. For collecting the diffraction data, the sample was contained in a thin-walled vanadium can of about 3-4 mm diameter and about 5 cm long, and rotated about the long axis to average out any preferred orientations. The data were obtained in a $2\theta$ range of $5^\circ \leq \theta \leq 105^\circ$ (at $0.05^\circ$ intervals), using a PSD (with 5 detector elements) that covered a range of $20^\circ$ at a time. Neutrons of wavelength 1.4875 Å were used for the diffraction experiments. Magnetization measurements were performed using a superconducting quantum interference device (SQUID) magnetometer (Quantum Design) in the temperature range 1.7-300 K and an applied field of 50 Oe.

## 3. Results and discussion

X-ray powder diffraction studies were performed on series of compounds $(Nd_{1-x}Ca_x)(Ba_{1.6}La_{0.4})Cu_3O_z$ with $x = 0.0$ to 0.6, showed them to be single phase. A fit of a typical neutron diffraction pattern is shown in Figure 1. Figure 2 shows the typical zero-field-cooled (ZFC) and field cooled (FC) magnetization data of $(Nd_{1-x}Ca_x)(Ba_{1.6}La_{0.4})Cu_3Oz$ with $x = 0.0$, 0.1, 0.2 and 0.4 with an applied filed of 50 Oe. In both ZFC and FC states data were acquired during warming of the sample. The superconducting transition temperature ($T_c$) is defined as the onset of diamagnetic response. It is seen from figure 2 and figure 5 that $T_c$ increases with increasing Ca concentration, which is in agreement with an earlier report [11]. For the sample with $x = 0.1$, the diamagnetic signal is small, suggesting that superconductivity is not a bulk property of the sample. It is evident from the figure that the Meissner fraction increases with increasing Ca content, which may indicate that a superconducting phase is forming while more $Nd^{3+}$ is replaced by $Ca^{2+}$. However, neither XRD nor neutron diffraction shows any evidence of another phase at room temperature.

The neutron diffraction data were analyzed by Rietveld refinement procedure using the Generalized Structural Analysis System (GSAS) program. The structure refinement carried out with orthrorhombic structure, space group *Pmmm*. It was realized that lattice parameter '*a*' and '*b*' are nearly the same, exhibit possible formation of tetragonal phase. Therefore, neutron diffraction data on all samples were refined on the basis of a tetragonal structure (space group *P4/mmm*). Neutron scattering factors used were (in units of fm) 8.27 for La, 4.90 for Ca, 5.25 for Ba, 7.69 for Nd, 7.718 for Cu and 5.805 for O. Structural parameters such as atomic coordinates, fractional occupancies and the thermal parameter (Ui) for different atoms (including various oxygen sites assigned to the standard R-123 structure) obtained from Rietveld analysis, are listed in Table 1.

In R-123 type compounds, the unit cell is considered as an assembly of three $ABO_3$ type pervoskite cells stacked one on top of the other with the central atom being R, in the middle cell and Ba in other two. This would result in a chemical formula $RBa_2Cu_3O_9$, and would contain a three-dimensional network of Cu-O bonds. The lower dimensionality and oxygen content $O_{7-\delta}$ arise from the missing oxygen atoms surrounding the R-site, Cu-O chains in the basal planes. The R and the two Ba atoms occupy crystallographically distinct sites with 8-fold and 10 fold oxygen coordination, respectively. There are two different types of Cu atoms, labeled as Cu(1) (0,0,0) for the Cu in the basal plane and Cu(2) (0,0,$z$) for the Cu in the two Cu-O planes. The Cu(1) atoms have strong covalent bonds to four $O_2$ atoms in a square planner configuration, while Cu(2) atoms are bonded to five oxygens in pyramidal symmetry. However, Cu-$O_2$ planes extend indefinitely in two direction (in the *ab*-plane), while Cu-O chains extend indefinitely in only one direction (the b-axis direction). The oxygen sites in the Cu-$O_2$ planes are identified as O(2) (½, 0, z) and O(3) (0, ½, z). The oxygen site in the Ba-O plane is designated as O(4) (0,0,z), and is often called the "apical oxygen". The oxygen sites in basal plane, often called as Cu-O chains, are designated as O(1)(0,½,0) {along *b*-axis} and O(5)(½,0,0) {along *a*-axis}. In orthorhombic R-123, the O(1) sites are fully occupied, while the O(5) sites are nearly vacant, giving rise to b > a. The O-T transition occur either by removing O(1) [oxygen deficient R-123 with oxygen stoichiometry < 7.0] or by filling the O(5) sites [e.g. trivalent La at Ba site, resulting in oxygen stoichiometry > 7.0]. [12,13]. Moreover, an interesting situation can occur via redistribution of O(1) and O(5) with oxygen stoichiometry close to 7.0, which is the unique case for the present case of $(Nd_{1-x}Ca_x)(Ba_{1.6}La_{0.4})Cu_3O_z$. In the tetragonal system with $a = b$ [Cu(1)-O(1) = Cu(1)-O(5)], the O(1) and O(5) are indistinguishable, and the same is true for O(2) and O(3) sites in the Cu-$O_2$ planes.

The observed variations of lattice parameter and cell volume with Ca content are shown in Figure 3. It is observed that the lattice parameter '*a*' and the cell volume both decrease up to $x = 0.4$. The decrease in lattice parameter reflects the decrease in Cu(2)–O(2) distance, and indicates a decrease in flatness of the Cu-$O_2$ plane with the decrease in "buckling angle" [Cu(2)–O(2)–Cu(2)], which helps to revive $T_c$ from 37 K ($x = 0.0$) to 84 K ($x = 0.4$).

Initially structure refinement model assumed with Nd/R and Ba/La ratios fixed by initial stoichiometry, but site occupancies were allowed to vary. In the subsequent analysis of neutron diffraction data, it was assumed that Nd and Ba fully occupy their normal sites. After fixing Nd and Ba occupancies, the Ca and La occupancies were allowed to vary freely both at the actual sites (Ca at Nd site and La at Ba site) and the nominal sites (La at Nd site and Ca at Ba site) with the condition that combined fractional and nominal at R site (Ca at Nd + La at Nd site) and Ba site (La at Ba + Ca at Ba) add up to unity (i.e. full occupancy). In order to maintain a meaningful and reliable variation of Ca and La occupancies at the two sites this procedure was adopted uniformly for all the samples. The variation of occupancies for La and Ca at different sites is shown in Fig. 4(a) and 4(b) respectively. The intermixing behaviour of La and Ca occupancy variation with R (R=Nd, Dy, and Y) through neutron diffraction has also been observed in La-1113, La-2125 and R-123 systems [20-23].

It is well known that Ca substitution at the R-site decreases the oxygen content of R-123 compounds; however, La substitution at the Ba-site increases the oxygen content [12,13]. Analysis of the neutron diffraction data reveals that some Ca atoms occupy the actual Nd-site while some occupy the Ba site. Consequently, corresponding amount of La substitutes at the Nd site. The oxygen content, obtained from refinement of neutron data, is found to be close to 7.0 for all Ca-doped $(Nd_{1-x}Ca_x)(Ba_{1.6}La_{0.4})Cu_3O_z$ compounds with $x = 0.0$ to 0.6. The cross substitution of La at the Nd site and Ca at the Ba site randomizes the chain site oxygen's bringing the O(1) and O(5) site occupancies of oxygen in the basal planes to same level, and forces the compound to crystallize in a *P4/mmm* tetragonal structure.

Here it is worth mentioning that the substitution of $Ca^{2+}$ for $Nd^{3+}$ can increase the number of mobile holes provided the oxygen content remains nearly constant. An increased number of holes oxidizes the $Cu-O_2$ layer. The oxidation of $Cu-O_2$ layer removes electrons from the $x^2-y^2$ orbitals, which have antibonding character in the in-plane Cu-O bonds. Therefore, as number of holes ($n_H$) increases, the in-plane Cu(2)-O(2) bond length ($r_{Cu-O}$) is shortened [16,17]. The effective Cu valence, estimated from the overall oxygen content ($z$) obtained from neutron diffraction data are listed in Table-1, which increases with increasing

Ca concentration. Simultaneous increase in $T_c$ with Cu valence is observed, and showed maximum $T_c$ of 84 K at $x = 0.4$ (Figure 5). The introduction of Ca into $(Nd_{1-x}Ca_x)(Ba_{1.6}La_{0.4})Cu_3O_z$ alters the normal and superconducting properties substantially [11]. The ionic radii of $Ca^{2+}$ in six-fold and eight-fold coordination are significantly different. The ionic size of $Nd^{3+}$ in eight-fold coordination is (0.995 Å), which matches more closely with that of $Ca^{2+}$ in six-fold (0.99 Å) than that of eight-fold coordination (1.12 Å). The decrease in lattice parameter $a$ and cell volume up to $x = 0.4$, indicates that Ca at the Nd site would have preferred a six fold coordination. The lattice parameter '$c$' remains nearly unchanged which is in good agreement with earlier reports of the possibility of six-fold coordination in those samples [18,19] with $Ca^{2+}$ substitution at R site in fully oxygen annealed Y-123 systems. As Ca content increases in the regime $x > 0.4$, some $Ca^{2+}$ might change its coordination from six-fold to eight-fold, which occupy Ba site and corresponding amount of La occupy at Nd site (Fig. 4(a) and 4(b); Table 1). Therefore, lattice parameter '$a$' and cell volume increases due to the fact that $La^{3+}$ and $Ca^{2+}$ in eight-fold coordination are bigger than Nd, which indicate an increase of the interplane Cu(2)-O(2) bond distances, and suggest a decrease in the number of holes in Cu-O$_2$ planes, which also results in an increase in the "buckling angle". As the buckling angle increases, the Cu-O$_2$ planes become flatter which results in suppression of superconductivity.

Magnetization and neutron diffraction results may be summarized as follows:

1. The superconducting transition temperature $T_c$ increases with increasing Ca content up to $x = 0.4$.

2. With progressive Ca doping at the Nd site, the lattice parameter $a$ decreases, but $c$ remains nearly unchanged, suggesting that $Ca^{2+}$ may adopt six-fold coordination.

3. $Ca^{2+}$ may change its coordination from six-fold to eight fold with simultaneous increases in both lattice parameter and cell volume for further increase in Ca content $x > 0.4$.

It is evident from Figure 5 that $T_c$ increases with increasing $x$ up to $x = 0.4$ and thereafter, decreases slightly for $x > 0.4$. The most spectacular proof that Ca doping enhances superconductivity is the phase $(Nd_{0.6}Ca_{0.4})(Ba_{1.6}La_{0.4})Cu_3O_z$, which exhibits a $T_c$ of 84 K, where as the Ca-free cuprate $Nd(Ba_{1.6}La_{0.4})Cu_3O_z$ does not superconduct down to 10 K and shows $T_c^{on} = 37$ K [11]. The substitution of Ca for Nd introduces oxygen vacancies and nonstoichiometry, which lead to hole doping. Therefore, this implies that the reduction in $T_c$ from 91 K for pristine Nd-123 to 37 K for $Nd(Ba_{1.6}La_{0.4})Cu_3O_z$ is compensated by appropriate hole doping with Ca. Hence, the oxide displaying the optimum $T_c \approx 84$ K at $x = 0.4$ is identified as a compensated oxide whose $T_c$ lies close to that of pristine Nd-123 ($T_c = 91$ K). Further increase in $x$ from 0.4 to 0.6 causes $T_c$ to decrease from 84 K ($x = 0.4$) to 78.5 K ($x = 0.6$) due to excess hole doping from Ca.

## 4. Conclusions

We conclude that Ca doping enhances superconductivity for doping level up to $x = 0.4$; thereafter it shows a decrease in $T_c$ with increasing $x$, due to excess hole doping. Neutron diffraction data suggest that there is intermixing of Ca and La occupancies. At higher doping level $x > 0.4$ indicates possibility of eight-fold coordination of Ca.

Table :I Lattice parameters, Cell volume, fractional occupancies, positional and thermal parameter for $(Nd_{1-x}Ca_x)(Ba_{1.6}La_{0.4})Cu_3O_z$ system obtained from Rietveld refinements of the neutron diffraction data. For the tetragonal refinements, O(1) and O(5) are equivalent, and O(2) and O(3) are equivalent. Numbers in parentheses are statistical standard deviations of the last significant digits.

| Parameter | $x = 0.0$ | $x = 0.1$ | $x = 0.2$ | $x = 0.3$ | $x = 0.4$ | $x = 0.5$ | $x = 0.6$ |
|---|---|---|---|---|---|---|---|
| Space group | P4/mmm | P4/mmm | P4/mmm | P4/mmm | P4/mmm | P4/mmm | P4/mmm |
| $a$ (Å) | 3.89505(24) | 3.88827(24) | 3.88239(22) | 3.87895(24) | 3.87681(26) | 3.88249(29) | 3.8823(4) |
| $b$ (Å) | 3.89505(24) | 3.88827(24) | 3.88239(22) | 3.87895(24) | 3.87681(26) | 3.88249(29) | 3.8823(4) |
| $c$ (Å) | 11.6897(8) | 11.7196(8) | 11.7212(7) | 11.7160(7) | 11.7157(8) | 11.7254(9) | 11.7205(14) |
| V (Å³) | 177.349(19) | 177.184(19) | 176.673(17) | 176.281(19) | 176.083(21) | 176.746(23) | 176.66(4) |
| **Nd (½,½,½)** | | | | | | | |
| $Ui \times 100$ (Å²) | 0.65(9) | 0.67(13) | 0.58(11) | 1.14(13) | 0.74(14) | 1.03(17) | 1.02(29) |
| $N_{Nd}$ | 1.0000 | 0.9000 | 0.8000 | 0.7000 | 0.6000 | 0.5000 | 0.4000 |
| $N_{Ca}$ | | 0.083(23) | 0.188(20) | 0.243(21) | 0.299(24) | 0.308(28) | 0.387(44) |
| $N_{La}$ | 0.0000 | 0.017(23) | 0.012(20) | 0.057(21) | 0.101(24) | 0.192(28) | 0.213(44) |
| **Ba (½,½,z)** | | | | | | | |
| Z | 0.18268(30) | 0.18561(28) | 0.18776(24) | 0.18709(25) | 0.18691(27) | 0.18677(33) | 0.1859(6) |
| $Ui \times 100$ (Å²) | 1.57(9) | 1.49(11) | 1.56(10) | 1.57(10) | 1.23(11) | 0.73(13) | 0.57(20) |
| $N_{Ba}$ | 1.6 | 1.6 | 1.6 | 1.6 | 1.6 | 1.6 | 1.6 |
| $N_{La}$ | 0.4 | 0.384(24) | 0.388(20) | 0.342(22) | 0.298(24) | 0.208(28) | 0.188(44) |
| $N_{Ca}$ | | 0.016(24) | 0.012(20) | 0.058(22) | 0.102(24) | 0.192(28) | 0.212(44) |
| **Cu(1) (0,0,0)** | | | | | | | |
| $Ui \times 100$ (Å²) | 1.87(10) | 2.04(10) | 2.17(9) | 2.46(10) | 2.41(11) | 2.35(13) | 3.82(27) |
| $N$ | 1.0000 | 1.0000 | 1.0000 | 1.0000 | 1.0000 | 1.0000 | 1.0000 |
| **Cu(2) (0,0,z)** | | | | | | | |
| Z | 0.34950(22) | 0.35093(22) | 0.35180(19) | 0.35214(21) | 0.35173(22) | 0.35193(27) | 0.3512(4) |
| $Ui \times 100$ (Å²) | 0.74(6) | 0.82(5) | 0.75(5) | 0.90(5) | 0.66(6) | 0.59(7) | 0.53(11) |
| $N$ | 2.0000 | 2.0000 | 2.0000 | 2.0000 | 2.0000 | 2.0000 | 2.0000 |
| **O(1) (0,½,0)** | | | | | | | |
| $Ui \times 100$ (Å²) | 3.54(28) | 4.08(32) | 4.72(31) | 5.23(31) | 4.43(33) | 4.67(38) | 2.62(54) |
| $N$ | 0.533(10) | 0.475(10) | 0.456(9) | 0.493(10) | 0.474(11) | 0.498(13) | 0.466(18) |
| **O(2) (½,0,z)** | | | | | | | |
| Z | 0.37003(20) | 0.36993(19) | 0.36971(16) | 0.36851(17) | 0.36882(18) | 0.36897(21) | 0.36820(34) |
| $Ui \times 100$ (Å²) | 1.15(6) | 1.24(6) | 1.25(5) | 1.44(5) | 1.23(6) | 1.02(7) | 1.13(11) |
| $N$ | 4.0000 | 4.0000 | 4.0000 | 4.0000 | 4.0000 | 4.0000 | 4.0000 |
| **O(4) (0,0,z)** | | | | | | | |
| Z | 0.15666(40) | 0.15522(43) | 0.15494(38) | 0.15427(41) | 0.15433(43) | 0.15395(56) | 0.15284(11) |
| $Ui \times 100$ (Å²) | 3.61(15) | 3.66(14) | 3.80(12) | 3.98(12) | 3.51(13) | 3.93(16) | 4.91(29) |
| $N$ | 2.0000 | 2.0000 | 2.0000 | 2.0000 | 2.0000 | 2.0000 | 2.0000 |
| **O(5) (½,0,0)** | | | | | | | |
| $Ui \times 100$ (Å²) | 3.54(28) | 4.08(32) | 4.72(31) | 5.23(31) | 4.43(33) | 4.67(38) | 2.62(54) |
| $N$ | 0.533(10) | 0.475(10) | 0.456(9) | 0.493(10) | 0.474(11) | 0.498(13) | 0.466(18) |
| Total oxygen content/formula unit | 7.066(20) | 6.950(20) | 6.912(18) | 6.986(20) | 6.948(22) | 6.996(26) | 6.932(36) |
| Cu valence | 2.244 | 2.2 | 2.208 | 2.2906 | 2.2986 | 2.364 | 2.35467 |

**Figure Captions**

Fig. 1   Typical ovserved (dots) and the calculated (solid curve) neutron diffraction patterns for $(Nd_{1-x}Ca_x)(Ba_{1.6}La_{0.4})Cu_3O_z$ samples with $x$ = 0.0, 0.1, 0.2 and 0.4 at room temperature. The difference pattern for each sample is shown at the bottom.

Fig. 2   Field-cooled and zero-field-cooled magnetization of $(Nd_{1-x}Ca_x)(Ba_{1.6}La_{0.4})Cu_3O_z$ samples with $x$ = 0.0, 0.1, 0.2, 0.3, 0.4, 0.5 and 0.6 as a function of temperature in an applied filed of 50 Oe showing superconducting transition temperatrue.

Fig. 3   Variation of lattice parameter and Cell volume with Ca concentration.

Fig. 4   Intermixing of occupancies with Ca concentration.

     (a) Variation of the La occupancy  on Nd and Ba site with Ca concentration.

     (b) Variation of the Ca occupancy at Nd and Ba site with Ca concentration.

Fig. 5   Variation of transition temperature $T_c$ with Ca concentration

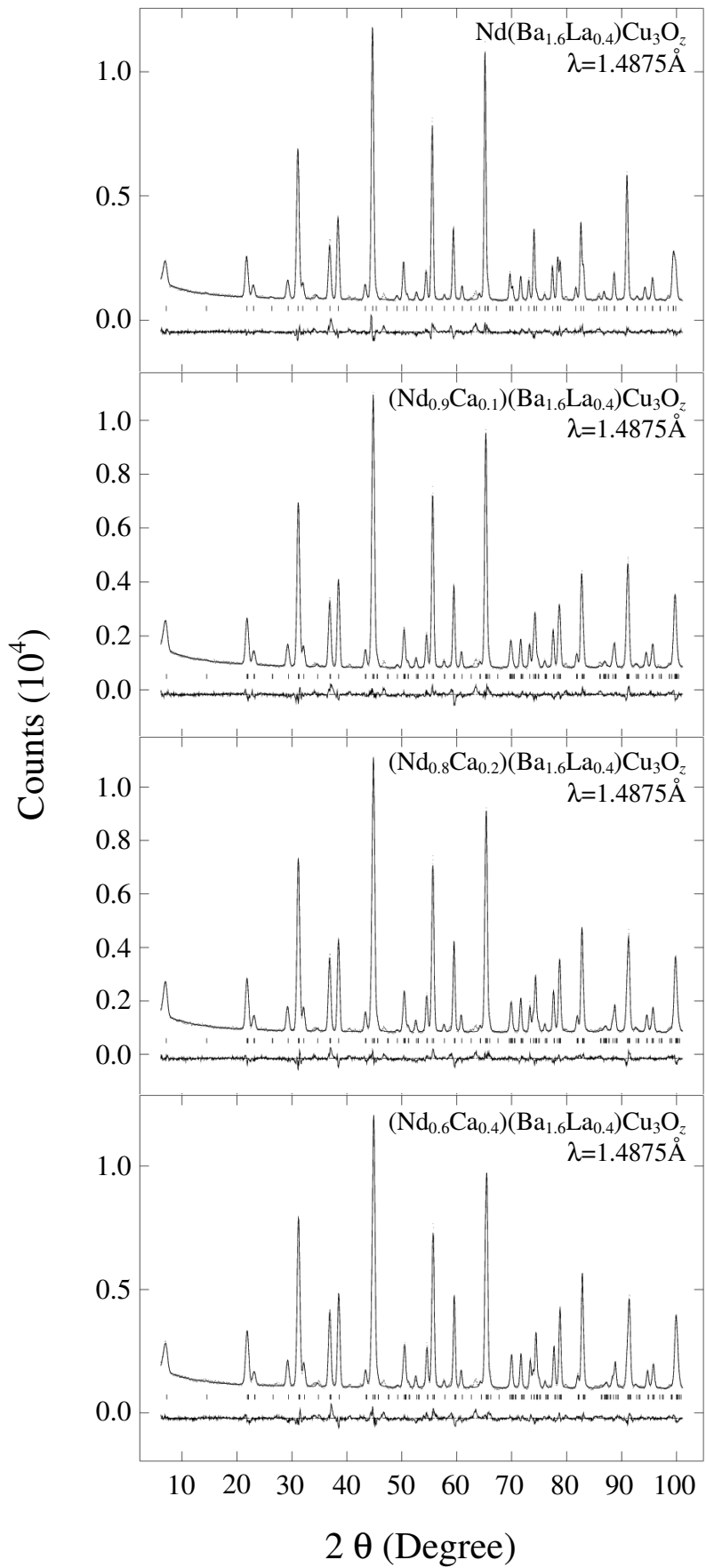

Fig. 1.

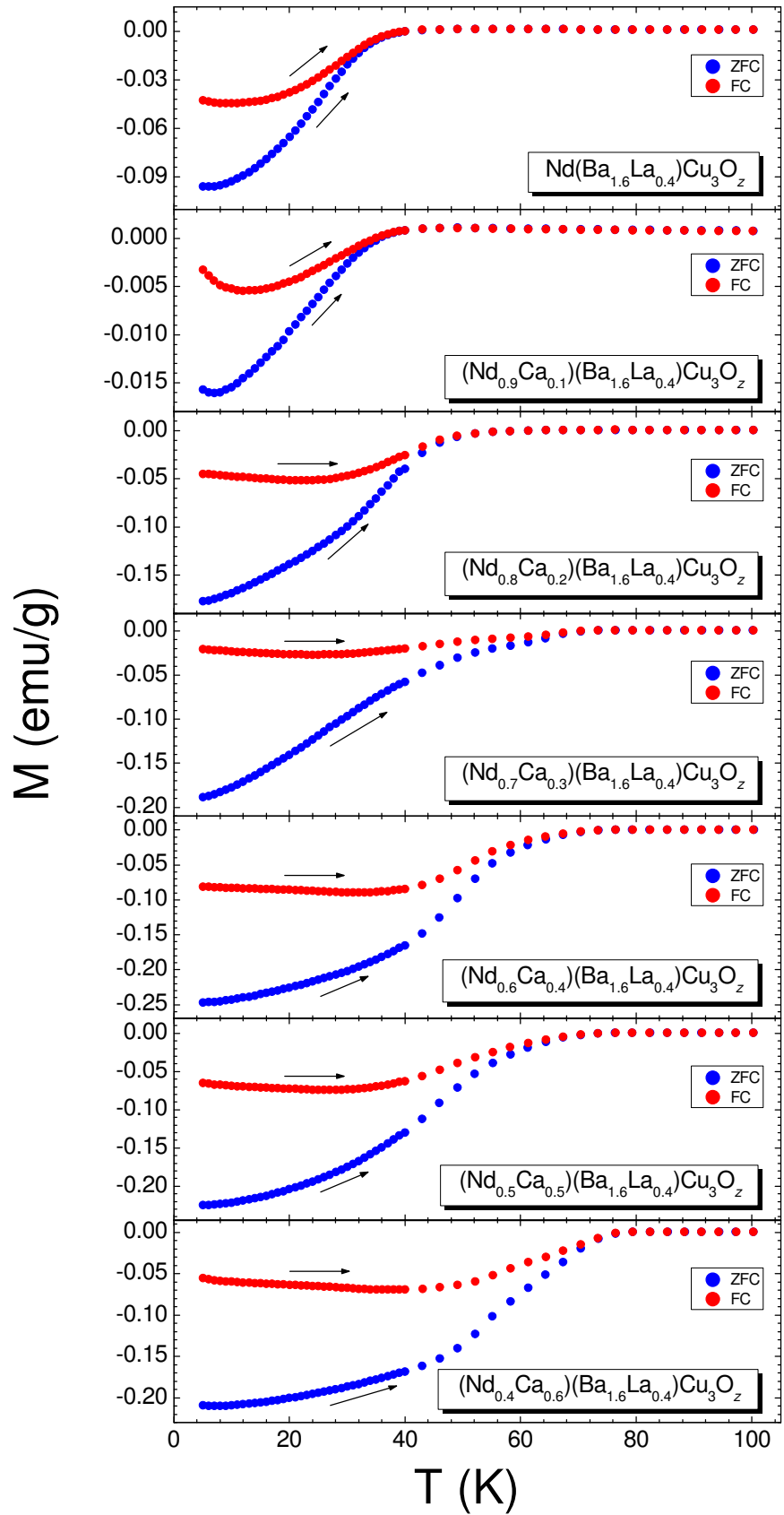

Fig. 2

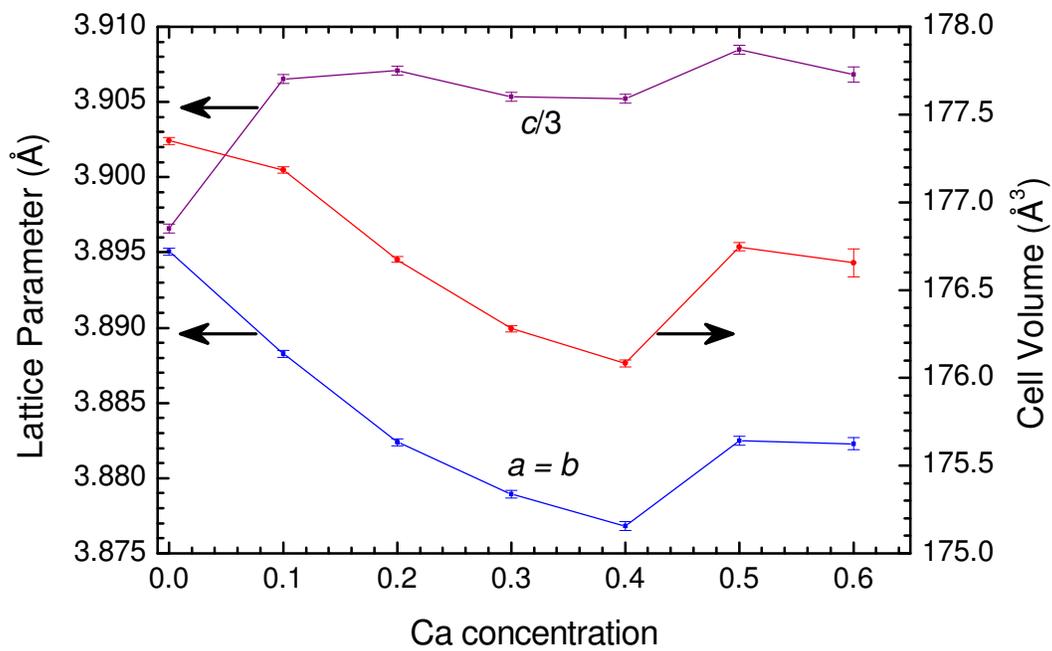

Fig. 3.

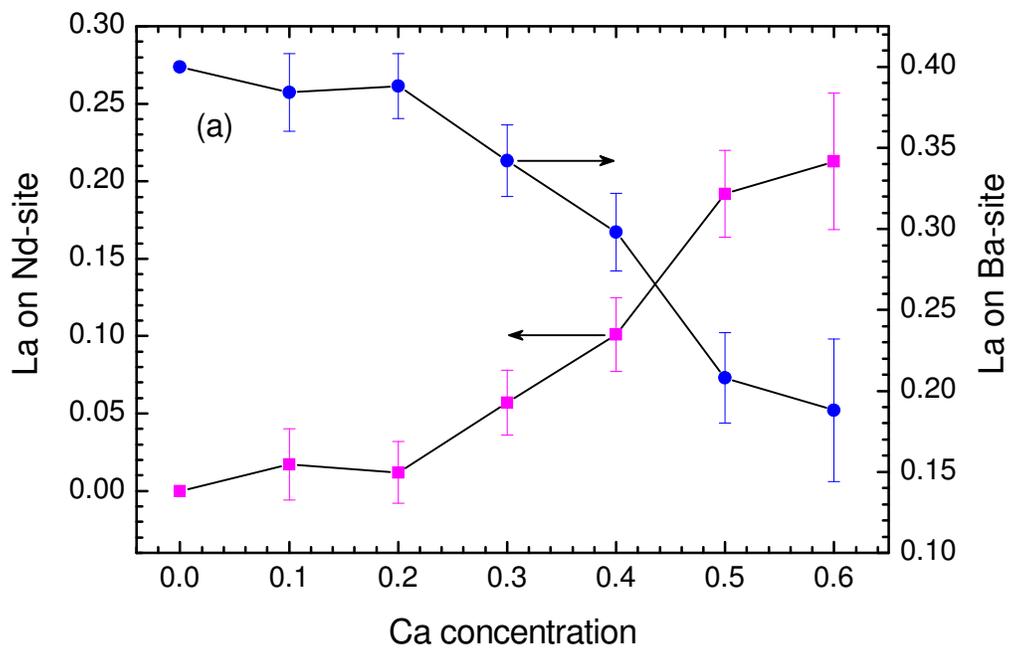

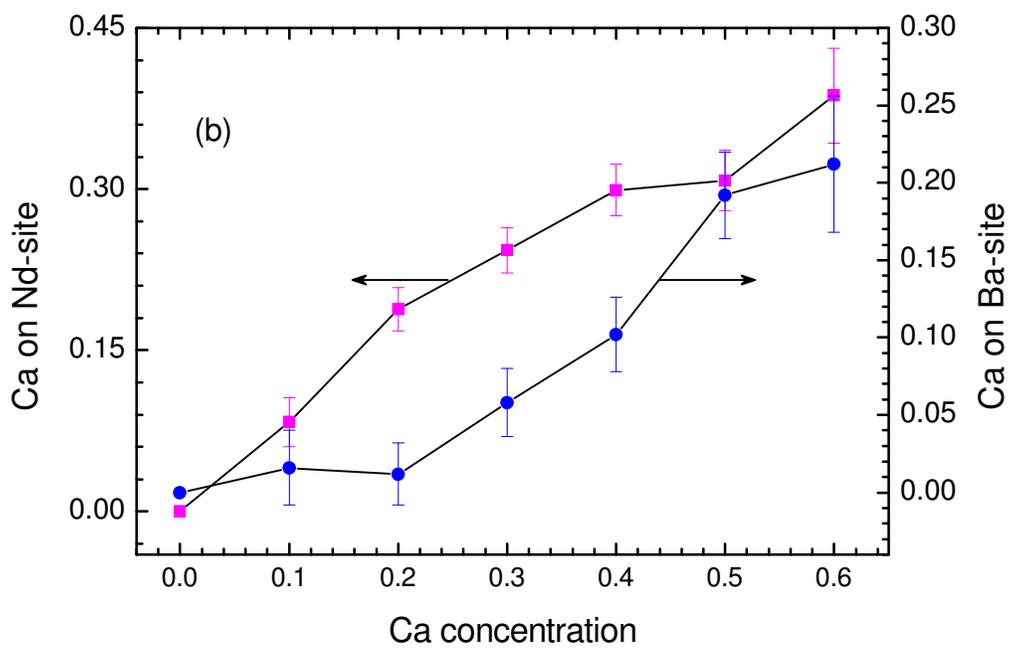

Fig. 4

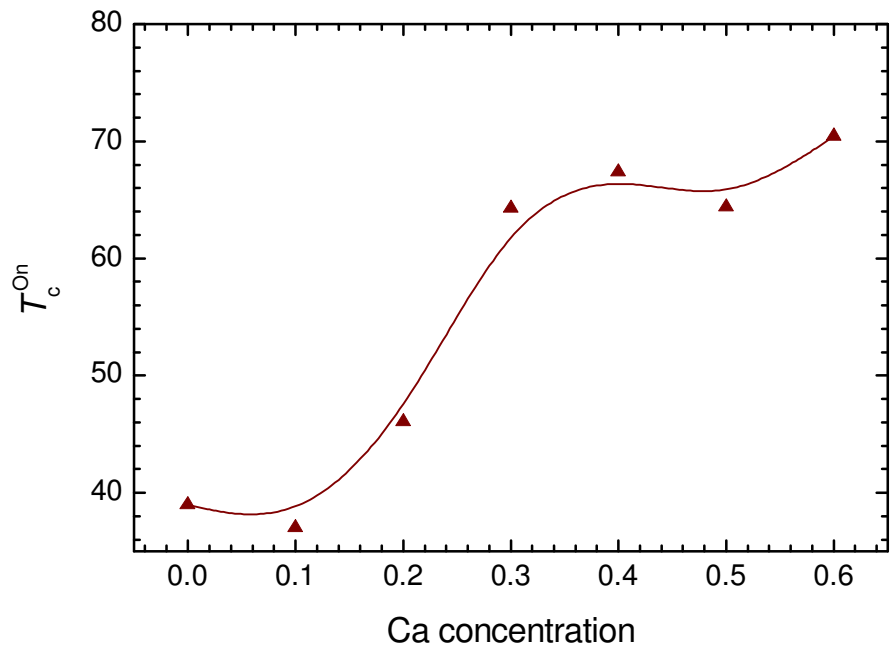

Fig. 5.